\documentclass[oneside,reqno,a4paper,12pt]{amsart}

\usepackage{latexsym}
\usepackage{euscript}
\usepackage{epsfig}

\date{\today}   
\renewcommand{\baselinestretch}{1.3} \large\normalsize   
\newcommand{\be}{\begin{equation}}
\newcommand{\ee}{\end{equation}}

\def\ie{{\it i.e.}}
\def\LEP2{{LEPII}}

\def\npb#1#2#3{    {\it Nucl. Phys. }{\bf B #1} (19#2) #3}

\def\plb#1#2#3{    {\it Phys. Lett. }{\bf B #1} (19#2) #3}
\def\prd#1#2#3{    {\it Phys. Rev. }{\bf D #1} (19#2) #3}

\begin{document}
\begin{titlepage}  
\begin{flushright}
BA-99-35
\end{flushright}
\vskip 1.5cm     
\title[]{Examining the cancellation mechanism of neutron \\ \vskip
0.2cm
EDM in a model with dilaton-dominated susy breaking}

\maketitle

\begin{center}
\textsc{S.M. Barr$^1$ and Shaaban Khalil$^{1,2}$} \\ \vspace*{2mm}
\small{\textit{$^1$Bartol Research Institute, University of
Delaware Newark, DE 19716}} \\ \vspace*{2mm}
\small{\textit{$^2$Ain Shams University, Faculty of Science, Cairo
11566, Egypt}} \\
\end{center}
\vspace*{2cm}
\begin{center}
ABSTRACT
\end{center}
\vspace*{8mm}
\begin{quotation}
We examine the cancellation mechanism between the different
contributions to the electric dipole moment of the neutron in a
model with dilaton-dominated SUSY breaking. We find these
accidental cancellations occur at few points in parameter space.
For a wide region of this space we must constrain the phase of
$\mu$ to be of order $10^{-1}$ and have the phases of $A$ and
$\mu$ strongly correlated in order to have small neutron EDM.
Moreover, we consider the indirect CP violation parameter
$\varepsilon$ in this region where the electric dipole moment is
less than the experimental limit and find that we can generate
$\varepsilon$ of order $10^{-6}$.
\end{quotation} 
\setcounter{page}{1}
\end{titlepage}     
\renewcommand{\baselinestretch}{1.3} \large\normalsize
\vspace*{4mm}
\section{Introduction}
        It is well known that in supersymmetric theories there are
new possibilities  for CP violation. In particular, the soft SUSY
breaking terms contain several parameters that  may be complex, as
can also be the $\mu$-parameter. These phases can cause at one
loop level an electric dipole moment (EDM) for the quarks and
leptons, and therefore also for the neutron. It has been known for
a long time that in a generic SUSY model the contributions to the
neutron electric dipole moment are larger than the experimental
limit $1.1\times 10^{-25}$ e cm unless either the new `SUSY
phases' are tuned to be of order $10^{-3}$, or the SUSY masses are
of order a TeV. Recently it has been suggested that a natural
cancellation mechanism exists whereby the electric dipole moment
of the neutron may be made small without such fine-tuning. In this
paper we examine this possibility in the context of a concrete
model of  SUSY breaking, namely dilaton-dominated breaking. \vskip
0.3cm In the minimal supersymmetric standard model there are
really only two new CP-violating phases. This can be seen as
follows. The parameters $M, A$ and $B$ and $\mu$ can be complex.
But of these four phases only two are physical. First, by an
R-rotation with R-charge $Q_R=1$ for lepton and quark superfields
and $Q_R=0$ for the vector and the Higgs superfields, the gaugino
mass parameter $M$ can be made real. Second, $B \mu$ can be made
real by a change of phase of the Higgs superfield. This ensures
that the Higgs vacuum expectation values are real. The remaining
phases cannot be defined away and violate CP. One is in $A=A_0
e^{i \phi_A}$ and the other in $B=B_0 e^{i\phi_B}$. The $\mu$
parameter then has a fixed phase $\mu=\mu_0 e^{-i\phi_B}$. In any
phase convention $$ \phi_A= \mathrm{arg}(AM^*)  \hspace{2cm}
\phi_B= \mathrm{arg}(BM^*).$$ The fact that there are two phases
can also be seen as follows. In the absence of the above soft SUSY
breaking terms and the bilinear $\mu$-term of the superpotential
there are two additional global $U(1)$ symmetries~\cite{nire} an
R-symmetry and a Peccei-Quinn symmetry. One finds that there are
three independent combinations of the four parameters which are
invariant under both $U(1)$'s, but only two of their phases are
independent, and can be chosen to be $\phi_A$ and $\phi_B$. \vskip
0.3cm It is these phases which must be of order $10^{-3}$ or less
if the neutron EDM  is not to be too large and the SUSY masses are
not unnaturally heavy. The common choice is to assume that
$\phi_A$ and $\phi_B$ are identically zero. However, unless there
is a symmetry which implies that these phases vanish or are very
small at the unification scale, it would be unnatural to assume
that they are. Clearly the SUSY phases problem is a problem of
SUSY breaking since the relevant phases originate from SUSY
breaking terms. \vskip 0.3cm
        The new suggestion for suppressing  the EDM of the neutron is made
in Ref.~\cite{nath1} where chromoelectric and purely gluonic
operator contributions are taken into account. It is proposed that
there are internal cancellations among the various components
contributing to the EDM which allows for the existence of large CP
violating phases. However the EDM analysis for the minimal
supergravity model~\cite{nath2} given in that paper actually shows
that there is only a tiny region in the parameter space where
these cancellations occur. Moreover, a heavy SUSY spectrum, of
order TeV and small SUSY phases of order $10^{-1}$ are considered
there. In fact, at these values the electric dipole moment is
already suppressed without any cancellation. The region where
large cancellations occur is bigger in Ref.~\cite{kane} where a
generic supersymmetric standard model is considered although all
the superpartner masses are kept light. The reason for this is
that there are more parameters to be adjusted to give a large
cancellation between the different EDM contributions. \vskip 0.3cm
We argue that this is an accidental cancellation and is as much a
fine tuning as the previously known mechanisms for suppression the
neutron EDM that use small phases or a heavy SUSY spectrum. In
this paper we examine this cancellation mechanism in a
string-inspired model where the supersymmetry is broken in the
hidden sector by the vacuum expectation values of the dilaton
and/or moduli fields. The soft SUSY breaking terms of this
scenario are very constrained and they are given in terms of two
parameters only. Indeed, we find that such a constrained model
does not allow for a large cancellation, the electric contribution
to EDM is the dominant one, and the only way to suppress it is
either by making the phase small or making the masses of order a
TeV. \vskip 0.3cm The paper is organized as follows. In section 2
we review the formulae of  the soft SUSY breaking terms in the
dilaton-dominated SUSY breaking model. In section 3 we examine the
cancellation of the electric dipole moment contributions. It is
found that this constrained model allows for this accidental
cancellation only at few points in the parameter space. Section 4
is devoted to studying the indirect CP violation parameter
$\varepsilon$ in the region where the electric dipole moment is
smaller than $10^{-25}$e cm. Finally, we give our conclusions in
section 5.
\section{dilaton dominated SUSY breaking}

        We consider an example of the string inspired model where the dilaton $S$
and overall modulus field $T$ contribute to SUSY breaking and the
vacuum energy vanishes. In this scenario, the soft masses $m_i$
and the gaugino masses $M_a$ are written as~\cite{munoz1}
\begin{eqnarray}
m^2_i &=& m^2_{3/2}(1 + n_i \cos^2\theta),
\label{scalar}\\
M_a &=& \sqrt{3} m_{3/2} \sin\theta e^{- i \alpha_{S}},
\label{gaugino}
\end{eqnarray}
where $m_{3/2}$ is the gravitino mass,
$n_i$ is the modular weight of the chiral multiplet and $\sin \theta$
corresponds to a ratio between the $F$-terms of $S$ and $T$.
For example, the limit, $\sin \theta \rightarrow 1$, corresponds to
the dilaton-dominated SUSY breaking.
Here the phase $\alpha_S$ originates from the $F$-term of $S$.
Similarly the $A$-parameters are also written as
\begin{eqnarray}
A_{ijk} &=& - \sqrt{3} m_{3/2} \sin\theta e^{-i \alpha_s}
- m_{3/2} \cos\theta
(3 + n_i + n_j + n_k) e^{-i \alpha_T},
\label{trilinear}
\end{eqnarray}
where $n_i$, $n_j$ and $n_k$ are the modular weights of the fields
that are coupled by this $A$ term. One needs a correction term in
eq~(\ref{trilinear}) when the corresponding Yukawa couplings
depend on moduli fields. This correction depends on the derivative
of the Yukawa couplings with respect to the moduli field, and is
therefore small since  the Yukawa couplings are constants or tend
exponentially to constants~\cite{vafa}. So we ignore them. Here
the phase $\alpha_T$ originates from the $F$-term of $T$. \vskip
0.3cm Finally, the magnitude of the scalar bilinear soft breaking
term $B \mu H_1 H_2$ depends on how the $\mu$-term is generated.
Therefore here we take $\mu$ and $B$ as free parameters and we fix
them by requiring successful electroweak symmetry breaking. \vskip
0.3cm Thus, gaugino masses and $A$-terms as well as the $B$-term
are, in general, complex. We have a degree of freedom to rotate
$M_a$ and $A_{ijk}$ at the same time~\cite{dugan}. Here we use the
basis where $M_a$ is real. Similarly we rotate the phase of $B$ so
that $B\mu$ itself is real. In this basis, the phases of $B$ and
$\mu$ satisfy $\phi_B = -\phi_{\mu} = {\rm arg}(BM^*)$. In
$A$-terms in the above basis, there remains only one independent
phase, namely, $\alpha' \equiv \alpha_T - \alpha_S$. i.e.,
$A_{ijk}=A_{ijk}(\alpha')$. We assume the following modular
weights for quark and lepton fields $$n_Q=n_U=n_{H_1}= -1$$ and
$$n_D=n_L=n_E=n_{H_2}=-2.$$ As will be seen later, this assumption
is favorable for electroweak breaking. Under this assumption we
have $A_t=A_b=A$, where $A$ is given by
\begin{equation}
A= - \sqrt{3} m_{3/2} \sin\theta + m_{3/2} \cos\theta e^{-i
\alpha'}. \label{Aterm}
\end{equation}
\vskip 0.3cm Given the boundary conditions in
eqs.~(\ref{scalar}-\ref{trilinear}) at the compactification scale,
we determine the evolution of the couplings and the mass
parameters according to their one loop renormalization group
equation in order to estimate the mass spectrum of the SUSY
particles at the weak scale. The radiative electroweak symmetry
breaking imposes the following conditions on the renormalized
quantities:
\begin{equation}
m^2_{H_1} + m^2_{H_2} + 2\mu^2 > 2 B \mu,
\end{equation}
\begin{equation}
(m^2_{H_1}+\mu^2)( m^2_{H_2} + \mu^2) < (B \mu)^2,
\end{equation}
\begin{equation}
\mu^2= \frac{m^2_{H_1} - m^2_{H_2}\tan^2\beta}{\tan^2\beta -1} -
\frac{M_Z^2}{2},
\end{equation}
and
\begin{equation}
\sin 2\beta = \frac{- 2 B \mu}{m^2_{H_1} + m^2_{H_2} + 2\mu^2},
\end{equation}
where $\tan\beta=\langle H_2^0 \rangle/\langle H_1^0 \rangle$ is
the ratio of the two Higgs VEVs that gives masses to the up and
down type quarks and $m_{H_1}$, $m_{H_2}$ are the two soft Higgs
masses at the electroweak scale. We take here $\tan\beta=3$, using
the above equations we can determine $\vert \mu \vert$ and $B$ in
terms of $m_{3/2}$, $\theta$ and $\alpha'$. The phase of $\mu$
($\phi_{\mu}$) remains undetermined.
\section{The Calculation of Electric Dipole Moment}
In supersymmetric theories, the EDM of the quark receives
contributions at the one loop level from diagrams in which the
charginos, neutralinos or gluinos are exchanged together with the
squarks. The EDM operator changes the chirality of the quark. The
gaugino couples the quark to the squark with the same chirality
via the gauge interactions, while the Higgsino couples the quark
to the squark with the opposite chirality via the Yukawa
interactions. In a supersymmetric model, this can happen in two
ways, either through the gaugino-Higgsino mixing or through the
mixing of $\tilde{q}_L$ and $\tilde{q}_R$. The chargino loop
diagrams involve both possibilities, while gluino-loop diagram
involves the second one only. \vskip 0.3cm Moreover, besides the
quark electric dipole, there are additional operators that
contribute to the EDM. They are the gluonic operator
$\mathcal{O}_G=-\frac{1}{6} f^{abc} G_a G_b \tilde{G}_c$ and the
quark chromoelectric dipole moment $\mathcal{O}_q=\frac{1}{4}
\bar{q}\sigma_{\mu \nu} \gamma_5 T^a G^{\mu \nu a}$, where $T^a$
are the generators of $SU(3)$. In Ref.~\cite{arnowit}, it was
estimated that the gluonic operator is the smallest contribution
when all the mass scales are taken to be equal. However, in
Ref.~\cite{nath2}, it was claimed that the contributions of the
chromoelectric and the purely gluonic operators can be comparable
to the contribution of the electric dipole operator. The
calculation of the gluino contributions to the quark EDM in that
reference shows that they are given by
\begin{equation}
{d_{q-gluino}^E}/{e}=-\frac{2\alpha_s}{3\pi} \sum_{k=1}^{2}
     {\rm Im}(\Gamma_{q}^{1k}) \frac{m_{\tilde{g}}}{M_{\tilde{q}_k}^2}
     Q_{\tilde{q}} {\rm B}(\frac{m_{\tilde{g}}^2}{M_{\tilde{q}_k}^2}),
\end{equation}
and
\begin{equation}
\tilde d_{q-gluino}^C=\frac{g_s\alpha_s}{4\pi} \sum_{k=1}^{2}
     {\rm Im}(\Gamma_{q}^{1k}) \frac{m_{\tilde{g}}}{M_{\tilde{q}_k}^2}
      {\rm C}(\frac{m_{\tilde{g}}^2}{M_{\tilde{q}_k}^2}).
\end{equation}
While the chargino loop contributions are given by
\begin{equation}
{d_{u-chargino}^{E}}/{e}=\frac{-\alpha_{\mathrm
EM}}{4\pi\sin\theta_W^2}\sum_{k=1}^{2}\sum_{i=1}^{2}
      {\mathrm Im}(\Gamma_{uik})
               \frac{\tilde{m}_{\chi_i}}{M_{\tilde{d}k}^2} [Q_{\tilde{d}}
                {\mathrm B}(\frac{\tilde{m}^2_{\chi_i}}{M_{\tilde{d}k}^2})+
        (Q_u-Q_{\tilde{d}}) {\mathrm
A}(\frac{\tilde{m}^2_{\chi_i}}{M_{\tilde{d}k}^2})],
\end{equation}

\begin{equation}
{d_{d-chargino}^{E}}/{e}=\frac{-\alpha_{\mathrm
EM}}{4\pi\sin\theta_W^2}\sum_{k=1}^{2}\sum_{i=1}^{2}
   {\mathrm Im}(\Gamma_{dik})
               \frac{\tilde{m}_{\chi_i}}{M_{\tilde{u}k}^2} [Q_{\tilde{u}}
                {\mathrm B}(\frac{\tilde{m}^2_{\chi_i}}{M_{\tilde{u}k}^2})+
        (Q_d-Q_{\tilde{u}}) {\mathrm
A}(\frac{\tilde{m}^2_{\chi_i}}{M_{\tilde{u}k}^2})],
\end{equation}
and
\begin{equation}
\tilde d_{q-chargino}^C=\frac{-g^2 g_s}{16\pi^2}\sum_{k=1}^{2}\sum_{i=1}^{2}
      {\mathrm Im}(\Gamma_{qik})
               \frac{\tilde{m}_{\chi_i}}{M_{\tilde{q}k}^2}
                {\rm B}(\frac{\tilde{m}^2_{\chi_i}}{M_{\tilde{q}k}^2}).
\end{equation}
Finally, the coefficient of the CP violating dimension six operator
is given by
\begin{equation}
d^G=-3\alpha_s m_t (\frac{g_s}{4\pi})^3 {\mathrm Im}
(\Gamma_{t}^{12})\frac{m^2_{\tilde{t}_1}-m^2_{\tilde{t}_2}}{m_{\tilde{g}}^5}
{\mathrm
H}(\frac{m^2_{\tilde{t}_1}}{m_{\tilde{g}}^2},\frac{m^2_{\tilde{t}_2}}{m_{\tilde{g}}^2},
\frac{m^2_t}{m_{\tilde{g}}^2}).
\end{equation}
In these equations, $M_{\tilde{q}_k}$ are the masses of the
corresponding scalar particle running in the loop. The functions
$\mathrm{A,B,C, H}$ and $\Gamma$ can be found in
Ref.~\cite{nath2}. Another contribution to the quark electric
dipole moment comes from the neutralino loops. However, as
explained in detail in Ref.~\cite{kane}, this contribution is
small and can be neglected in the electric dipole moment analysis.
There are several reasons for that. For instance, the elements of
the neutralino diagonalizing matrix yield smaller imaginary parts
than the elements of the chargino matrices. Also the chargino
contribution is enhanced due to large values of the loop function
compared to the function in the neutralino expression. \vskip
0.3cm The analysis of the EDM above is at the electro-weak scale
and it must be evolved down to the hadronic scale via the
renormalization group evolution as explained in Ref~\cite{nath2}.
The contribution to the neutron EDM coming from the $d^E$ is
obtained by using the $SU(6)$ quark model which gives $$d_n =
\frac{1}{3}(4 d_d -d_u).$$ The contributions to $d_n$ coming from
the $\tilde{d}^C$ and $d^G$ are estimated using naive dimensional
analysis~\cite{georgi}. As stated above, we determine $\mu$ and
$B$ from the electroweak breaking conditions. Hence the parameter
space of this model consists of only $m_{3/2}$, $\theta$, $\phi_A$
and $\phi_{\mu}$. In this scenario we take $\phi_{\mu}$ to be
unconstrained while $\phi_A$ is given in terms of $\alpha'$ at GUT
scale. It is worth mentioning that $\phi_A$ at electroweak scale
is less than $\alpha'$ since the real part of $A$ is running
faster than the imaginary part. In figure (1) we show the values
of the total EDM of the neutron as function of $m_{3/2}$, for
$\cos^2 \theta \simeq 1/2$ and $\phi_{\mu}=\alpha'=\pi/2$. \vskip
0.3cm
\begin{figure}[h]
\psfig{figure=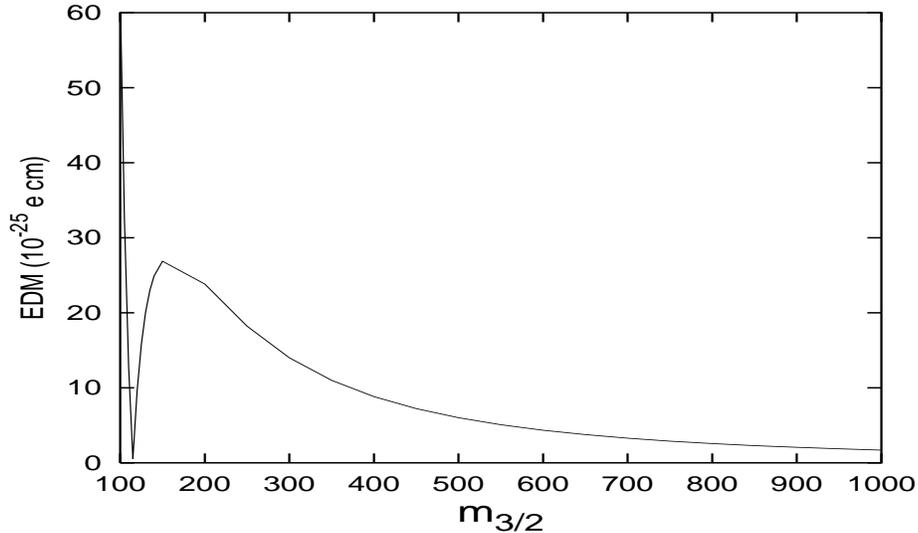,height=7cm,width=12cm} \caption{The total
EDM as function of $m_{3/2}$, for $\phi_{\mu}=\alpha'=\pi/2$.}
\end{figure}
\vskip 0.3cm From this figure we note that with these large phases
we still have to have $m_{3/2}>$ TeV to obtain value for the EDM
less than the experimental limit $1.1\times 10^{-25}$e cm. ({\it
i.e.}, all the scalar masses are of order TeV). The experimental
limit on the chargino mass puts a lower bound on $m_{3/2}$, namely
$m_{3/2} > 100$ GeV. In the region of $m_{3/2}$ between 100 and
115 GeV the chromoelectric contribution to the EDM exceeds the
electric contribution and at one point in the parameter space
these two contributions are approximately equal so that we find
the total EDM of neutron is less than the experimental limit.
However, as we have said, this is happening in a very tiny region
in the parameter space. Now, to see the effect of each of the
phases on the EDM values we plot in figure (2) the EDM versus the
$\phi_{\mu}$ for two values of $\alpha'$. The solid line
corresponds to the EDM where $\alpha'=0$, while the dashed line
corresponds to the EDM where $\alpha'$ is of order $\pi/2$. \vskip
0.3cm
\begin{figure}[h]
\psfig{figure=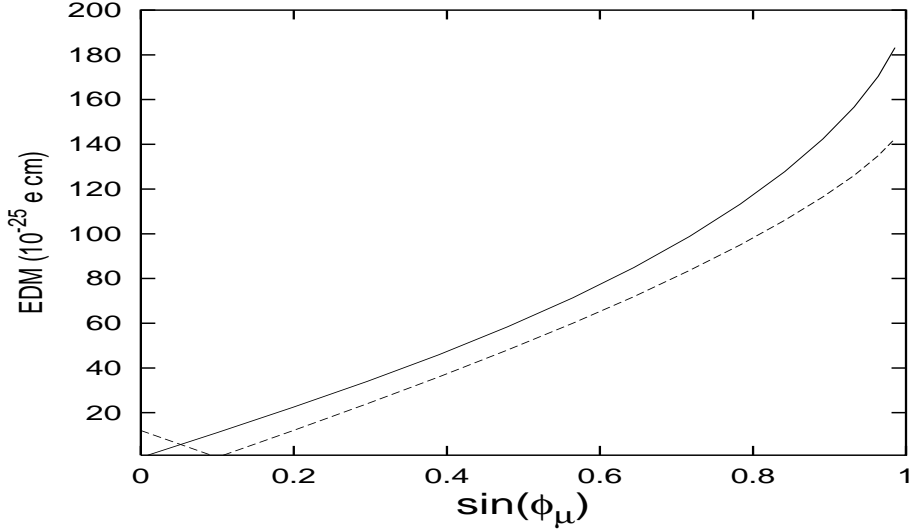,height=7cm,width=12cm} \caption{The total
EDM versus $\sin\phi_{\mu}$. The solid line corresponds
$\alpha'=0$ and the dashed one corresponds to $\alpha'=\pi/2$.}
\end{figure}
\vskip 0.3cm This figure shows that the values of the EDM of the
neutron in the case of non-vanishing  $\phi_A$ is less than in the
case of $\phi_A=0$. This is an interesting result since by
adjusting the phase of $\mu$ we can leave the phase of $A$ large
and have the  EDM of the neutron less than the experimental bound.
This could generate, as we examine in the next section, a sizable
value for the CP violation $\varepsilon$ parameter. For fixed
value of $m_{3/2}\simeq \mathcal{O}(100)$ GeV and $\cos^2\theta
\simeq 1/2$, the behaviour of the neutron EDM in the phase plane
$\phi_{\mu}-\alpha'$ can be understood as follows. The interesting
region of this plane (\ie, where the neutron EDM is less than the
experimental limit) is where the phases are opposite in sign and
are strongly correlated. For instance, the constraint obtained on
the phase $\phi_{\mu}$ is more severe for $\alpha'=0$ than the one
obtained for non vanishing $\alpha'$, as Figure (2) confirms.
\vskip 0.3cm In fact, in our model we have left the $B$ parameter
free and determined it at electroweak scale from the radiative
breaking conditions so that the phase of $\mu$ in this case is a
free parameter. However in Ref.~\cite{munoz1} three sources for
the $B$ parameter were considered, labeled by $B_Z$, $B_{\mu}$ and
$B_{\lambda}$. The source of $B_Z$ is the presence of certain
bilinear term in the K\"ahler potential which can naturally induce
a $\mu$-term of order $m_{3/2}$ after SUSY breaking. An
alternative mechanism to generate a $B$-term in a scalar potential
is to assume that superpotential $W$ includes a SUSY mass term
$\mu(S,T)H_1 H_2$ induced by a non-perturbative effect, then a
$B$-term is automatically generated and it is called $B_{\mu}$.
Also, it has been pointed out that the presence of a
non-renormalizable term in the superpotential $\lambda W H_1 H_2$
yields dynamically a $\mu$ parameter when $W$ acquires VEV. The
corresponding $B$-term is denoted by $B_{\lambda}$. Furthermore,
it was shown in Ref.~\cite{khalil1} that $B_{\mu}$ is the
favorable choice to realize the radiative electroweak symmetry
breaking. In mixed dilaton/modulus SUSY breaking the $B_{\mu}$ is
given by
\begin{equation}
B_{\mu} = m_{3/2} [- e^{i\alpha_S} -\sqrt{3} \sin\theta
-\sqrt{3}\cos \theta (3 +n_{H_1} + n_{H_2}) e^{-i\alpha'}],
\end{equation}
with the values of the modular weights we gave in section 2
this formula leads to
\begin{equation}
B_{\mu} = m_{3/2} [- e^{i\alpha_S} -\sqrt{3} \sin\theta],
\end{equation}
and therefore $$\tan(\phi_{\mu})= \frac{\sin
\alpha_S}{\cos\alpha_S + \sqrt{3} \sin\theta}.$$ Thus, the phase
of $\mu$ in this case is constrained such that $\vert
\phi_{\mu}\vert \leq \pi/4$ (from $m_{H_2}$ with the modular
weights we are assuming (see eq.(\ref{scalar})) one has
$\cos^2\theta < 1/2 $. This implies that $\sin \theta >0$ and even
close to one). In other cases of generating the $B$-term, as well
as for other choices for the modular weights, we could get a
stronger bound on the phase of $\mu$. So it may be natural to have
such a small phase of $\mu$. Some further comments are in order.
It is important to note that in this model a large value of
$\alpha'$ leads to a large value of $\vert A\vert$; and this is
required to reduce the EDM value, as was also found in
Ref.~\cite{nath2}. Moreover, the sign of the gluino contribution
is reversed for large value of $\alpha'$, and a destructive
interference between it and the chargino contribution occurs. This
can be seen by looking to the lowest approximation of the gluino
contribution to the electric dipole moment $${d_{q-gluino}^E}/{e}
\simeq \frac{2\alpha_s}{3\pi} m_{\tilde{g}} Q_{\tilde{q}}
\frac{m_q}{M_{\tilde{q}_k}^4}(\vert A_q \vert \sin \phi_{A_q} +
\vert \mu \vert \sin \phi_{\mu} R_q)
B(\frac{m^2_{\tilde{g}}}{M_{\tilde{q}_k}^2}).$$ Thus, for $\vert
\phi_{\mu} \vert < 10^{-1}$ and $\vert \alpha' \vert =\pi/2$, and
$\phi_{\mu}$ and $\alpha'$ having opposite signs, the sign of the
gluino contribution is opposite to the sign of the chargino
contribution, and consequently there is some cancellation at that
point. It is important to note that this will not be the case if
the phase of $\mu$ is larger than $10^{-1}$. Finally, with the
phase ${\alpha'}$ of order $\pi/2$ and the phase of ${\mathrm
\mu}$ of order $10^{-1}$ we find the limit on the EDM of the
neutron is satisfied for a large region of the parameter space. We
have to mention that the cancellation between the electric and
chromoelectric contributions helps but it is not the reason for
reaching this region. The reason is that the value of the electric
contribution to the EDM of the neutron by itself is small. \vskip
0.3cm
\begin{figure}[h]
\psfig{figure=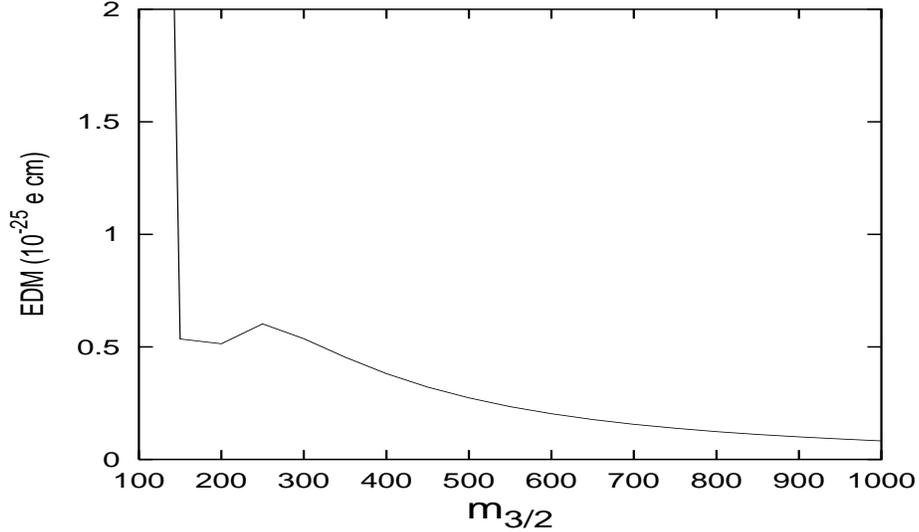,height=7cm,width=12cm} \caption{The total
EDM versus the gravitino mass where $\phi_{\mu}\simeq 10^{-1}$ and
$\alpha'=\pi/2$.}
\end{figure}
\vskip 0.3cm In this connection it is very important to mention
that in case of $\alpha'=\pi/2$, the phase of $A$ at GUT scale
from eq.(\ref{trilinear}) is of order $\pi/6$. This is the maximum
phase of the $A$-term in this model. Moreover, due to the running
from the GUT scale to weak scale this phase is reduced more and
becomes of order $\pi/20$, \ie, it is of the same order as
$\phi_{\mu}$. In fact, it is a feature of all supersymmetric
models that the phase of the trilinear couplings at the weak scale
is less than at the GUT scale due to the difference in the running
between the real and imaginary parts of $A$. On the other hand,
the phase of $\mu$ does not run, so that it has to be of order
$10^{-1}$ at the GUT scale too. It is the smallness of these
phases seperately, and not some cancellation between phases, that
is chiefly responsible for the smallness of the EDM. \vskip 0.3cm
We have not discussed the electron EDM, since the cancellation
mechanisms we have been considering (between gluino and chargino
diagrams, and between chromoelectric and electric dipole
operators) obviously do not apply to it. Even in the region of
parameter space where the neutron EDM is suppressed, there is no
reason to expect the electron EDM to be suppressed, and therefore
the electron EDM would remain a problem, especially as the
sleptons are lighter than the squarks.

\section{The parameter $\varepsilon$}
It is very important to show if it is possible to generate a
sufficiently large value of $\varepsilon$ in the region of the
parameter space that we found in the last section leads to values
for the EDM of the neutron less than the experimental limit. It is
not important whether the values of the phases at the large scale
or at the low scale are of order one or $10^{-1}$. What is really
important is that these phases alone can generate a sizable
contribution to the CP violation processes. \vskip 0.3cm The value
of the indirect CP violation in the Kaon decays, $\varepsilon$, is
defined as
\begin{equation}
\varepsilon = \frac{e^{i\frac{\pi}{4}}
{\rm Im} M_{12}}{\sqrt{2} \Delta m_K},
\end{equation}
where $\Delta m_K= 2 {\rm Re} \langle K^0 \vert H_{eff} \vert \bar{K}^0
\rangle = 3.52 \times 10^{-15}$ GeV.
The amplitude
$M_{12}=\langle K^0 \vert H_{eff} \vert \bar{K}^0 \rangle$.
The relevant supersymmetric contributions to $K^0-\bar{K}^0$ are
the gluino and the chargino contributions, (\ie, the transition
proceeds through box diagrams exchanging gluino-squarks and
chargino-squarks). It is usually expected that the gluino is the
dominant contribution. However, as we will show, it is impossible
in the case of degenerate $A$-terms that the gluino gives any significant
contribution to $\varepsilon$ when the CKM matrix is taken to be real even with
large phase of $A$.
\vskip 0.3cm
The amplitude of the gluino contribution is given in
Ref.~\cite{masiero} in terms of the mass insertion $\delta_{AB}$
defined by $\delta_{AB} = \frac{\Delta_{AB}}{\tilde{m}^2}$ where
$\tilde{m}$ is an average sfermion mass and $\Delta$ is off-diagonal
terms in the sfermion mass matrices. The mass insertion to accomplish
the transition from $\tilde{d}_{iL}$ to $\tilde{d}_{jL}$ ($i,j$ are
flavor indices) is given by~\cite{hagelin}
\begin{eqnarray}
(\Delta^d_{LL})_{ij}&\simeq&-\frac{1}{8\pi^2}\left[\frac{K^{\dag}
(M_u^{diag})^2 K}{v^2 \sin^2\beta} \ln(\frac{M_{GUT}}{M_W})
\right](3\tilde{m}^2+\vert X \vert^2),
\\
(\Delta^d_{LR})_{ij}&\simeq&-\frac{1}{8\pi^2}\left[\frac{K^{\dag}
(M_u^{diag})^2 K\ M_d}{v^2 \sin^2\beta \cos\beta}
\right] \ln(\frac{M_{GUT}}{M_W}) X,
\\
(\Delta^d_{RL})_{ij}&\simeq&-\frac{1}{8\pi^2}\left[\frac{M_d\ K^{\dag}
(M_u^{diag})^2 K }{v^2 \sin^2\beta \cos\beta}
\right] \ln(\frac{M_{GUT}}{M_W}) X,
\\
(\Delta^d_{RR})_{ij}&=&0,
\end{eqnarray}
where $X= A_d -\mu\ \tan\beta $. It is clear that $\Delta_{ij}$ in
general are complex due to the complexity of the CKM matrix, the
trilinear coupling $A$ and $\mu$ parameter. Here we assume the
vanishing of $\delta_{CKM}$ to analyze the effect of the SUSY
phases. We notice that $(\Delta^d_{LL})_{12}$ is proportional to
$\vert X \vert^2$ \ie, it is real and does not contribute to
$\varepsilon$ whatever the phase of $A$ is. Moreover, the values
of the $(\Delta^d_{LR})_{12}$ and $(\Delta^d_{RL})_{12}$ are
proportional to $m_s$ and $m_d$, hence they are quite small.
Indeed in this case we find the gluino contribution to
$\varepsilon$ is of order $10^{-6}$. The contribution is enhanced
in case of non-degenerate $A$-terms~\cite{khalil2}. \vskip 0.3cm
For the chargino contribution the amplitude is given
by~\cite{branco}
\begin{eqnarray}
\langle K^0 \vert H_{eff} \vert \bar{K}^0 \rangle &=& -\frac{G_F^2
M_W^2}{(2\pi)^2} (V_{td}^* V_{ts})^2 f_K^2 M_k [ \frac{1}{3}
C_1(\mu) B_1(\mu)
\nonumber\\
        &+& (\frac{M_k}{m_s(\mu) +m_d(\mu)})^2
(-\frac{5}{24} C_2(\mu) B_2(\mu) + \frac{1}{24} C_3(\mu)
B_3(\mu))].
\end{eqnarray}
The complete expression for these function can be found in
Ref.~\cite{branco}. Since we have $\tan \beta \simeq 3$ the value
of $C_3$ is much smaller than $C_1$ since it is suppressed by the
ratio of $m_s$ to $M_W$. However, by neglecting the flavor mixing 
in the squark mass matrix $C_1$ turned out to be exactly real~\cite{demir}. 
The imaginary part of $C_1$ is associated to the size of the 
intergenerational sfermion mixings, thus it is maximal for large 
$\tan \beta$. In low $\tan \beta$ case, that we consider, the imaginary part 
of $C_1$ is very small, and the gluino contribution is still the dominant 
contribution $\varepsilon$. As explained, the EDM of the
neutron constrains the phase of $\phi_{\mu}$ to be of order
$10^{-1}$, where $\alpha' \simeq \pi/2$ and $\cos^2 \theta \simeq
1/2$. In this region we estimate the value of $\varepsilon$. It
turns out that it is of order $10^{-6}$, which is less
than the experimental value $2.26\times 10^{-3}$. 

\section{Conclusions}
We have examined the cancellation mechanism between the different
contributions for the electric dipole moment of the neutron in a
model with dilaton-dominated SUSY breaking. We found that this is
an accidental cancellation and is as much a fine tuning as the
previously known mechanisms for the suppression the EDM of the
neutron. It occurs only at a few point in the parameter space of
this model. \vskip 0.3cm Furthermore, we studied the indirect CP
violation parameter $\varepsilon$ and we showed that it is of
order $10^{-6}$ only in the region where the EDM of the
neutron is smaller than $10^{-25}$e cm. \vskip0.75truecm
\begin{center}
\noindent{\small ACKNOWLEDGEMENTS}
\end{center}
\vskip0.5truecm S.K. would like to acknowledge the support given
by the Fulbright Commissionand the hospitality of the Bartol
Research Institute. S.M.B. is supported in part by the Department
of Energy under contract No. DE-FG02-91ER-40626. 
\end{document}